\documentclass[aps,pra,twocolumn,groupedaddress]{revtex4-2}

\usepackage{graphicx}
\usepackage{amsmath}

\begin{document}


\title{Observation antibunching with classical light in a linear interferometer}


%
%
%
\author{Yu Gu}
\thanks{These authors contributed equally to this work.}

\author{Yuhan Ma}
\thanks{These authors contributed equally to this work.}

\author{Yiqi Song}

\author{Meixue Chen}

\author{Hui Chen}

\author{Huaibin Zheng}

\author{Yuchen He}

\author{Zhuo Xu}

\author{Jianbin Liu}
\email[]{liujianbin@xjtu.edu.cn}
\affiliation{Electronic Materials Research Laboratory, Key Laboratory of the Ministry of Education \& International Center for Dielectric Research, School of Electronic Science and Engineering, Xi’an Jiaotong University, Xi'an 710049, China}

\author{Yu Zhou}

\author{Fuli Li}

\affiliation{MOE Key Laboratory for Nonequilibrium Synthesis and Modulation of Condensed Matter, Department of Applied Physics, Xi’an Jiaotong University, Xi’an, 710049, China}


\date{\today}

\begin{abstract}
Understanding the boundary between classical and nonclassical phenomena is important for both fundamental researches in quantum optics and applications in quantum information. One of the most interesting research  directions in this field is exploring nonclassical effects with classical light. In this paper, we will show that it is possible to observe antibunching with thermal light in a Hanbury Brown-Twiss interferometer by treating single-photon detectors as photon-number-resolving detectors to perform photon-number projection measurements. Both temporal and spatial antibunching is observed via the correlation of two detectors detecting one and zero photon, respectively. By comparing the measured results of thermal and laser light, it is found that the observed antibunching arises from the combined effect of photon statistics of thermal light and photon-number projection measurement.The classical and nonclassical nature of the observed antibunching is analyzed. The results are helpful to understand the connection between classical and nonclassical correlation and may find applications in multiphoton interference and quantum imaging.
\end{abstract}


\maketitle

\section{Introduction\label{intro}}

Since the beginning of quantum mechanics, the distinction between classical and quantum worlds has constituted a fundamental question in physics \cite{wheeler2014quantum}. Technologies based on quantum theory offer better precision \cite{Pan2012Multiphoton}, sensitivity \cite{lloyd2008enhanced}, security \cite{nielsen2010quantum}, along with other advantages. However, quantum resources typically suffer from drawbacks such as difficulty in generation \cite{anwar2021entangled}, low efficiency \cite{Lounis2005}, and susceptibility to degradation in environmental noise \cite{zurek2003decoherence}. It is tempting to employ classical resources to achieve similar priorities as the one with quantum resources with suitable modifications. For instance, secure quantum communication requires single-photon source, which suffers from low generating rates and low efficiency \cite{Lounis2005}. Hence weak pulsed laser light is usually employed as the light source in quantum communication with the help of post-selection and other technologies \cite{Scarani2004}. There are other studies focusing on simulating quantum systems with classical systems \cite{peng2014delayed,Qian2015Shifting,altmann2018quantum}.  In this paper, we will show that antibunching can also be observed with classical light in a linear interferometer.

Photon antibunching seems like the opposite effect of two-photon bunching of thermal light  \cite{brown1956correlation},  in which photons tend to arrive at detectors individually instead of in bunches.  However, the physical origins of these two effects are different. Two-photon bunching of thermal light is a result of two-photon interference \cite{shih2020introduction}. Antibunching is the reflection of corpuscular properties of photons and can not be interpreted in classical theory \cite{paul1982photon,loudon2000quantum}. Based on the criterion given by Glauber and Surdarshan that a phenomenon can only be interpreted in quantum theory is nonclassical \cite{glauber1963quantum,sudarshan1963equivalence}, antibunching is a nonclassical effect.  Photon antibunching was first observed by Kimble \textit{et al.} in resonance fluorescence of atoms \cite{kimble1977photon}. Later, antibunching was also reported with entangled photon pairs \cite{grangier1986experimental, nogueira2001experimental}, Cavity QED \cite{hennrich2005transition}, two-atom system \cite{wolf2020light}, quantum dot \cite{hanschke2020origin},  microwave-frequency photons \cite{bozyigit2011antibunching}, free fermions \cite{kiesel2002observation,iannuzzi2006direct,rom2006free}, and other nonclassical systems. As long as photons or particles are in number state (mainly single-particle state), antibunching can be observed in a Hanbury Brown-Twiss (HBT) interferometer \cite{loudon2000quantum}.

Even though photon antibunching is known as a nonclassical effect, antibunching was reported with classical light in a HBT interferometer via post-selection \cite{wu2011two,luo2012nonlocal,meyers2012positive}. Antibunching with classical light was also reported in a modified HBT interferometer \cite{ye2022antibunching} or when these two detectors are not at the symmetrical positions \cite{kondakci2016hanbury}. The reported antibunching \cite{wu2011two,luo2012nonlocal,meyers2012positive, ye2022antibunching, kondakci2016hanbury} can be interpreted in classical theory and hence a classical effect. Later, it was shown that antibunching can be observed with classical light by employing photon-number-resolving detectors (PNRDs) in a HBT interferometer  \cite{cao2024quantum,dawkins2024quantum,you2024isolating,mostafavi2025multiphoton}. Different from the observed antibunching with classical light in Refs. \cite{wu2011two,luo2012nonlocal,meyers2012positive}, the observed antibunching with PNRDs can not be interpreted in classical theory \cite{cao2024quantum,dawkins2024quantum,you2024isolating,mostafavi2025multiphoton}, which indicates that the observed antibunching may be a nonclassical effect. Different theories were employed to interpret the observed antibunching. Cao \textit{et al.} employed probability theory to interpret their results and mainly focused on computational ghost imaging with digital micromirror device (DMD) \cite{cao2024quantum}. Mostafavi \textit{et al.} employed similar method to study computational ghost imaging with PNRDs \cite{mostafavi2025multiphoton}. You \textit{et al.} employed DMD and spatial light modulator to study the correlation of photons with different angular momenta \cite{you2024isolating}. Dawkings \textit{et al.} employed quantum Gaussian–Schell model to study the spatial correlation of multiphoton detection events of thermal light \cite{dawkins2024quantum}. However, detailed discussions about the origin of antibunching in their systems are missing \cite{cao2024quantum,dawkins2024quantum,you2024isolating,mostafavi2025multiphoton}. Recently, we reported a ghost imaging experiment with thermal light based on antibunching, where a negative ghost image can be obtained by post-selecting on zero-photon events. Nevertheless, the origin of antibunching in this context remains to be fully understood \cite{chen2025ghost}. In this paper, we will provide a complementary framework that unifies these observations  \cite{cao2024quantum,dawkins2024quantum,you2024isolating,mostafavi2025multiphoton,chen2025ghost} into a comprehensive understanding. Besides spatial antibunching, temporal antibunching is also reported.  A direct comparison of laser and thermal light in the same experimental framework demonstrates that the observed antibunching is jointly determined by the photon-number projection measurement and photon statistics of thermal light. The results clarify the physics of the observed antibunching in thermal light and may find potential applications in multiphoton interference and quantum imaging \cite{cao2024quantum,dawkins2024quantum,chen2025ghost}.

\section{Theory}

\begin{figure}[htb]
\centering
\includegraphics[width=45mm]{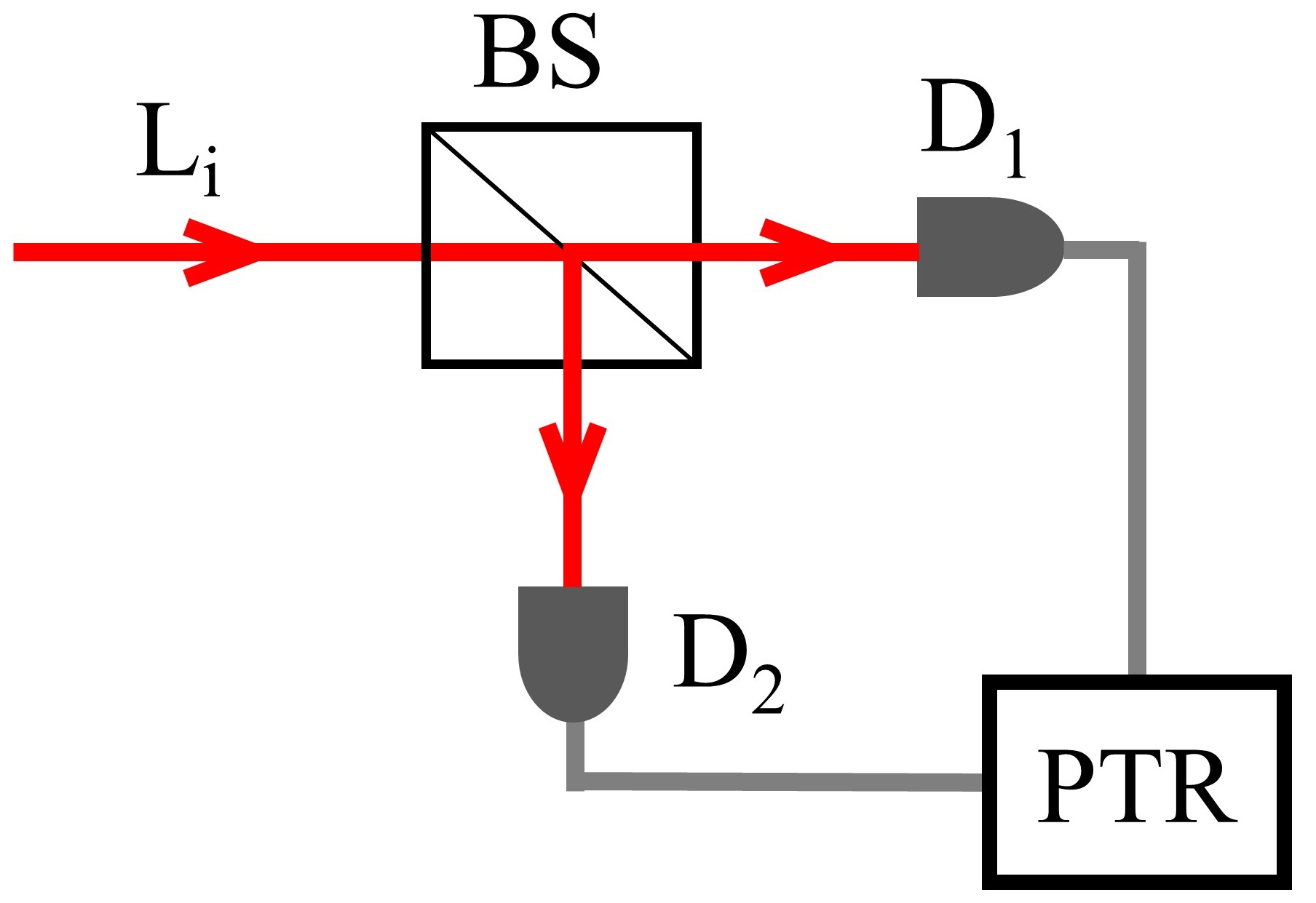}
\caption{Scheme for the modified HBT interferometer. L$_\text{i}$: Thermal (i=T) or laser (i=L) light. BS: Non-polarizing $1:1$ beam splitter. D$_1$ and D$_2$: Single-photon detectors. PTR: Photon detection time series recording system.}\label{1-scheme}
\end{figure}

Figure \ref{1-scheme} shows the scheme for measuring antibunching with classical light, which is the same as HBT interferometer except two-photon coincidence count measurement system (CC) is replaced by photon detection time series recording system (PTR). A light beam is split into two identical beams by a 1:1 non-polarizing beam splitter (BS). The input light beam can be either thermal (L$_\text{T}$) or laser light (L$_\text{L}$), or any other type of light. Two single-photon detectors (D$_1$ and D$_2$) are employed to record the photon arriving time information in these two light beams, respectively. Photon detection events from these two detectors are time-tagged independently using the same clock, allowing for off-line correlation analysis.

In our earlier research, we obtained the joint probability generating function of thermal light in the scheme shown in Fig. \ref{1-scheme} \cite{chen2025ghost},
\begin{eqnarray}\label{probability}
&&M(x,y; \vec{r}_1,t_1;\vec{r}_2,t_2)\\
&=&\frac{1}{1 - \bar{n}(x-1) - \bar{n}(y-1) + (1-\mu)\bar{n}^2 (x-1)(y-1)},  \nonumber
\end{eqnarray}
where $x$ and $y$ are auxiliary variables for probability generating function, $(\vec{r}_1,t_1)$ and $(\vec{r}_2,t_2)$ are the space-time coordinates of photon detection events at D$_1$ and D$_2$, respectively, $\bar{n}$ is the average number of photons detected by D$_1$ or D$_2$, $\mu=|g^{(1)}(\vec{r}_1,t_1;\vec{r}_2,t_2)|^2$, $g^{(1)}(\vec{r}_1,t_1;\vec{r}_2,t_2)$ is the normalized correlation function of light field at $(\vec{r}_1,t_1)$ and $(\vec{r}_2,t_2)$. The probability function of D$_1$ detecting $m$ photons and D$_2$ detecting $n$ photons can be obtained via the Taylor expansion coefficient of $M(x,y; \vec{r}_1,t_1;\vec{r}_2,t_2)$ when $x=y=0$ \cite{mandel1995optical,chattamvelli2019generating},
\begin{eqnarray}\label{probability-final}
&&P_{mn}(\vec{r}_1,t_1;\vec{r}_2,t_2) \nonumber\\
&=& \left. \frac{1}{m! \, n!} \frac{\partial^{m+n} M(x, y; \vec{r}_1,t_1;\vec{r}_2,t_2)}{\partial x^m \partial y^n} \right|_{x=0, y=0}.
\end{eqnarray}

Substituting Eq. (\ref{probability}) into Eq. (\ref{probability-final}), the probability function of D$_1$ detecting $m$ photons and D$_2$ detecting $n$ photons can be simplified as (see Appendix A)
\begin{eqnarray}\label{pmn-final}
&&P_{mn}(\vec{r}_1,t_1;\vec{r}_2,t_2) \nonumber\\ 
&=&\sum_{k=0}^{\min(m,n)}\frac{(m+n-k)!}{(m-k)!\,(n-k)!\,k!}\, \\
&&\times \frac{\bigl[\bar{n}+(1-\mu)\bar{n}^2\bigr]^{m+n-2k}\bigl[-(1-\mu)\bar{n}^2\bigr]^k}{\bigl[1+2\bar{n}+(1-\mu)\bar{n}^2\bigr]^{m+n-k+1}}. \nonumber
\end{eqnarray}

The normalized correlation function of D$_1$ detecting $m$ photons and D$_2$ detecting $n$ photons is defined as
\begin{eqnarray}\label{gmn-def}
g_{mn}^{(2)}(\vec{r}_1,t_1;\vec{r}_2,t_2) \equiv \frac{P_{mn}(\vec{r}_1,t_1;\vec{r}_2,t_2) }{P_{m}(\vec{r}_1,t_1)P_{n}(\vec{r}_2,t_2) },
\end{eqnarray}
where $P_{m}(\vec{r}_1,t_1)$ is the probability of D$_1$ detecting $m$ photons at $(\vec{r}_1,t_1)$, $P_{n}(\vec{r}_2,t_2)$ is the probability of D$_2$ detecting $n$ photons at $(\vec{r}_2,t_2)$. For thermal light, $P_{m}(\vec{r}_1,t_1)$ and $P_{n}(\vec{r}_2,t_2)$  is given by \cite{loudon2000quantum}
\begin{equation}\label{pm}
P_q = \frac{\bar{n}^q}{(1+\bar{n})^{q+1}}, 
\end{equation}
where $q=m$ or $n$, $\bar{n}$ is the average number of photons detected at $(\vec{r}_j,t_j)$ ($j=1$, 2) and $\bar{n}$ is independent of time.

Substituting Eqs. (\ref{pmn-final}) and (\ref{pm}) into Eq. (\ref{gmn-def}), the normalized correlation function of D$_1$ detecting $m$ photons and D$_2$ detecting $n$ photons can be obtained. The expression of $g_{mn}^{(2)}(\vec{r}_1,t_1;\vec{r}_2,t_2)$ is complicate for general $m$ and $n$. Two special cases are discussed in the following.

(I) $n=0$

There is only one term, $k=0$, left in the sum over $k$ in Eq. (\ref{pmn-final}) in this case. The normalized correlation function of D$_1$ detecting $m$ photons and D$_2$ detecting zero photons equals \cite{chen2025ghost}
\begin{eqnarray}
g_{m0}^{(2)}(\vec{r}_1,t_1;\vec{r}_2,t_2)&=& \frac{(1+\bar{n})^{m+2}\bigl[1+(1-\mu)\bar{n}\bigr]^m}{\bigl[1+2\bar{n}+(1-\mu)\bar{n}^2\bigr]^{m+1}}.
\end{eqnarray}

(II) $m=n=1$

There are two terms left in the sum over $k$ in Eq. (\ref{pmn-final}) in this case, which correspond to $k=0$ and $k=1$ terms. $P_{11}(\vec{r}_1,t_1;\vec{r}_2,t_2)$ can be simplified as (see Appendix A)
\begin{eqnarray}\label{p11}
&&P_{11}(\vec{r}_1,t_1;\vec{r}_2,t_2) \nonumber\\
&=& \frac{\bar{n}^{2}\bigl[(1+\mu)+2(1-\mu)\bar{n}+(1-\mu)^{2}\bar{n}^{2}\bigr]}{\bigl[1+2\bar{n}+(1-\mu)\bar{n}^{2}\bigr]^{3}}.
\end{eqnarray}
Substituting Eqs. (\ref{pm})  and (\ref{p11}) into Eq. (\ref{gmn-def}), it is straightforward to obtain
\begin{eqnarray}
&&g_{11}^{(2)} (\vec{r}_1,t_1;\vec{r}_2,t_2) \nonumber \\
&= &\frac{(1+\bar{n})^4\bigl[(1+\mu) + 2(1-\mu)\bar{n} + (1-\mu)^2\bar{n}^2\bigr]}{\bigl[1+2\bar{n}+(1-\mu)\bar{n}^2\bigr]^3},
\end{eqnarray}
where $\bar{n}$ and $\mu$ are functions of space-time coordinates, $\bar{n}(\vec{r}_1,t_1)$ is assumed to equal $\bar{n}(\vec{r}_2,t_2)$ in order to simplify the expression.

\begin{figure}[htb]
\centering
\includegraphics[width=75mm]{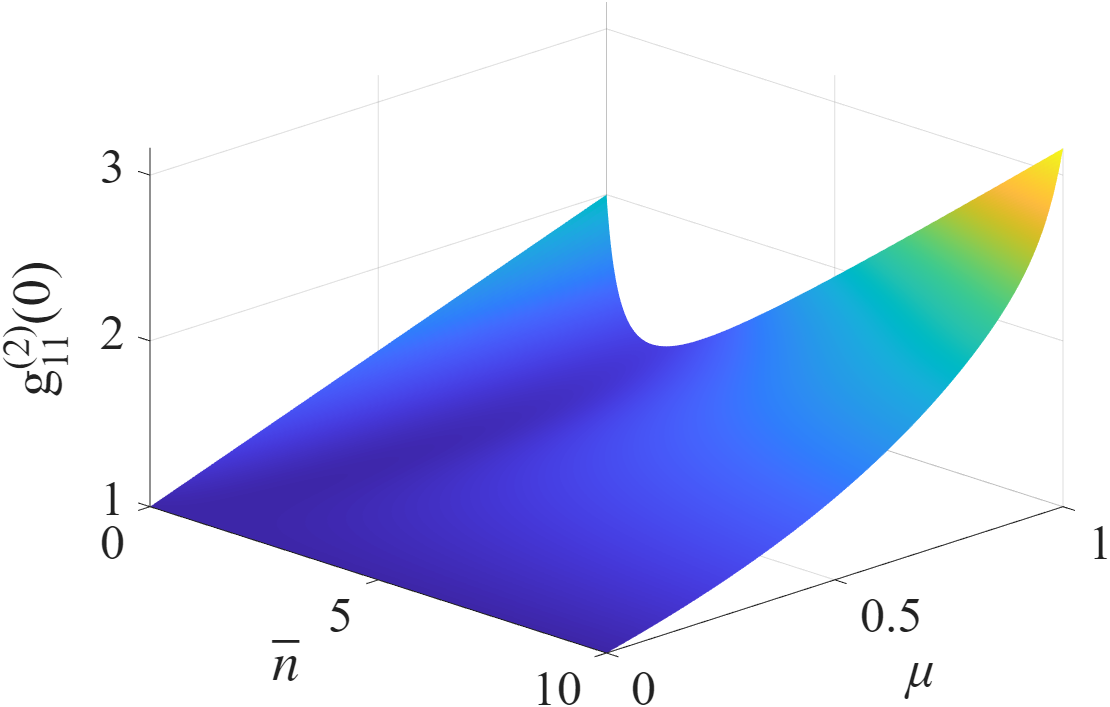}
\caption{The dependence of $g_{11}^{(2)} (0)$ on $\bar{n}$ and $\mu$. $g_{11}^{(2)} (0)$ is the normalized second-order correlation function when the two single-photon detection events at D$_1$ and D$_2$ are at the same space-time coordinate.  $\bar{n}$ is the average number of detected photons and $\mu$ is the squared modulus of the first-order coherence function of thermal light.}\label{2-g11}
\end{figure}

In order to analyze the relationship between $g_{11}^{(2)} (\vec{r}_1,t_1;\vec{r}_2,t_2)$ and the parameters, $\bar{n}$ and $\mu$, Fig.  \ref{2-g11} shows $g_{11}^{(2)} (0)$ for different values of $\bar{n}$ and $\mu$, where  $g_{11}^{(2)} (0)$ is the normalized second-order correlation function when the two single-photon detection events at D$_1$ and D$_2$ are at the same space-time coordinate. Based on the simulation in Fig. \ref{2-g11}, $g_{11}^{(2)} (0)$ equals 1 for different value of $\bar{n}$ when $\mu=0$, which is consistent with fact that two photon detection events are independent when the employed thermal light is incoherent. For a fixed average number of photons, $\bar{n}$, $g_{11}^{(2)} (0)$ increases as $\mu$ increases, which indicates that correlation of photon detection events increase as the coherence of thermal light increases. On the other hand, $g_{11}^{(2)}(0)$ is not monotonic in $\bar{n}$ for a fixed value of $\mu$. For coherent thermal light, \textit{i.e.}, $\mu=1$, $g_{11}^{(2)}(0)$ first decreases as $\bar{n}$ increases from zero and obtains its minimum when $\bar{n}=1$. Then $g_{11}^{(2)}(0)$ increases as $\bar{n}$ increases from 1. Figure \ref{2-g11} shows that the normalized correlation function, $g_{11}^{(2)} (0)$, is not only dependent on the coherence properties of thermal light, but also on the average number of detected photons.

\begin{figure}[htb]
\centering
\includegraphics[width=75mm]{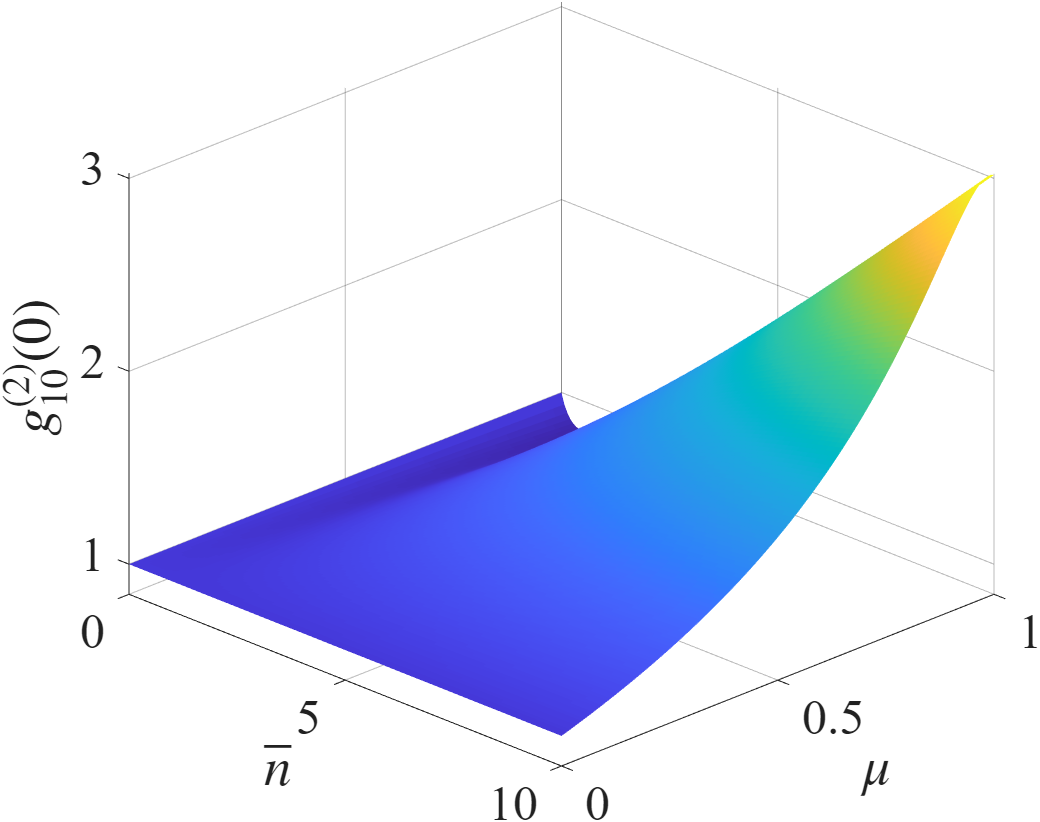}
\caption{The dependence of $g_{10}^{(2)} (0)$ on $\bar{n}$ and $\mu$. The meanings of the symbols are similar as the ones in Fig. \ref{2-g11}.}\label{3-g10}
\end{figure}

Figure \ref{3-g10} shows the dependence of $g_{10}^{(2)} (0)$ on $\bar{n}$ and $\mu$. Unlike $g_{11}^{(2)}(0)$,  $g_{10}^{(2)} (0)$ can fall below 1. For instance, $g_{10}^{(2)} (0)$ obtains its minimum, 0.84, when $\bar{n}=0.5$ and $\mu=1$, in which antibunching can be observed. Similar behaviors hold for $g_{m0}^{(2)} (0)$ when $m=2, 3, 4...$ except the minimum value of $g_{m0}^{(2)} (0)$ decreases as $m$ increases. $g_{00}^{(2)} (0)$ can not be less than 1, which indicates that antibunching can not be observed via $g_{00}^{(2)}$.

\begin{figure}[htb]
\centering
\includegraphics[width=90mm]{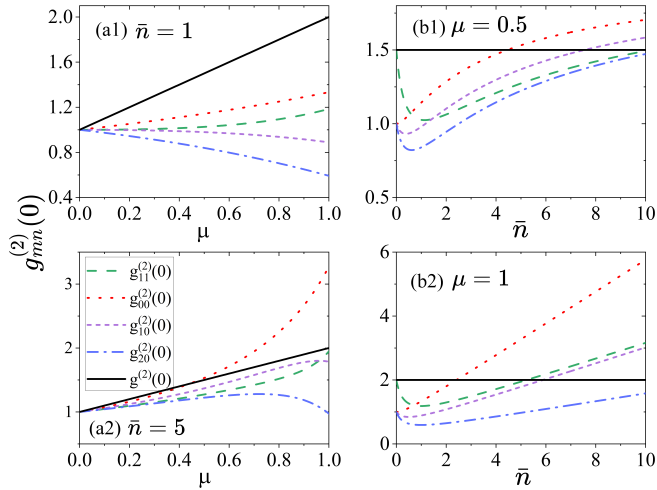}
\caption{The dependence of $g_{mn}^{(2)} (0)$ on $\bar{n}$ and $\mu$. (a1) and (a2) show the relation between $g_{mn}^{(2)} (0)$ and $\mu$ for $\bar{n}$ equaling 1 and 5, respectively. (b1) and (b2) show the relation between $g_{mn}^{(2)} (0)$ and $\bar{n}$ for $\mu$ equaling 0.5 and 1, respectively. $g^{(2)} (0)=1+\mu$ is the degree of second-order coherence of partially coherent thermal light and is drawn in these figures for comparison.}\label{4-gmn}
\end{figure}

In order to analyze the relationship between $g^{(2)}_{mn}$ and the parameters,  $\bar{n}$ and $\mu$, in detail, Fig. \ref{4-gmn} shows how $g_{mn}^{(2)} (0)$ varies for different $\bar{n}$ and $\mu$. $g^{(2)} (0)=1+\mu$ is the degree of second-order coherence of partially coherent thermal light and is employed for comparison \cite{loudon2000quantum}. Figures \ref{4-gmn}(a1) and (a2) show the dependence of $g_{mn}^{(2)} (0)$ on $\mu$ for $\bar{n}$ equaling 1 and 5, respectively. $g_{11}^{(2)} (0)$ and $g_{00}^{(2)} (0)$ increase as $\mu$ increases, which is a reflection of two-photon bunching of thermal light increases as $\mu$ increases.  In Fig. \ref{4-gmn}(a1), $g_{10}^{(2)} (0)$ and $g_{20}^{(2)} (0)$ decrease as $\mu$ increases and can fall below 1, which indicates that antibunching can be observed in this condition. While in  Fig. \ref{4-gmn}(a2), $g_{10}^{(2)} (0)$ and $g_{20}^{(2)} (0)$ first increase and then decrease as $\mu$ increases. The reason for the difference is that the value of $g_{mn}^{(2)} (0)$  is also dependent on the average number of detected photons, as shown in Figs. \ref{4-gmn}(b1) and (b2). For a fixed $\mu$, $g^{(2)} (0)$ of thermal light is independent of the average number of detected photons. $g_{00}^{(2)} (0)$ increases as $\bar{n}$ increases. $g_{10}^{(2)} (0)$, $g_{20}^{(2)} (0)$ and $g_{11}^{(2)} (0)$ first decrease and then increases as $\bar{n}$ increases. The minimum value of $g_{11}^{(2)} (0)$ can not be less than 1 while the minimum values of $g_{10}^{(2)} (0)$, $g_{20}^{(2)} (0)$ can fall below 1. It meas that antibunching can only observed via $g_{10}^{(2)}$ and $g_{20}^{(2)} $ in suitable conditions, but not via $g_{11}^{(2)} $ or $g_{00}^{(2)}$.

\section{Experiments}

\begin{figure}[htb]
\centering
\includegraphics[width=75mm]{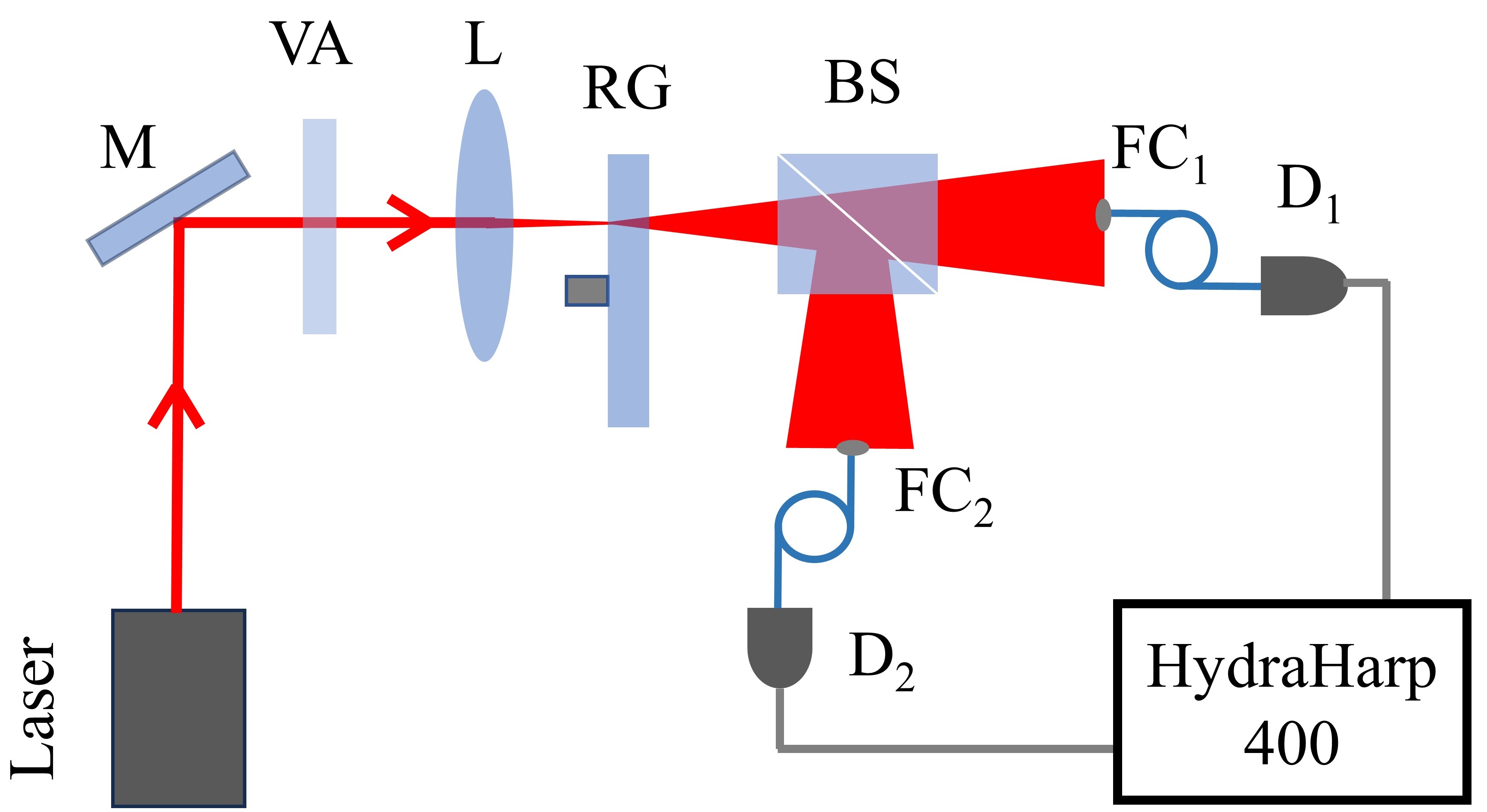}
\caption{Experimental setup to observe antibunching with classical light. Laser: Single-mode continuous-wave laser. VA: Variable light intensity attenuator. L: Lens. RG: Rotating groundglass. FC$_1$ and FC$_2$: Fiber couplers. HydraHarp 400: Photon detection time series recording system, HydraHarp 400, which can be replaced by other module with similar functions.}\label{5-setup}
\end{figure}

The experimental setup to verify the above predictions is shown in Fig. \ref{5-setup}. A single-mode continues-wave laser with central wavelength at 780 nm and frequency bandwidth of 200 kHz is employed as the light source. Pseudothermal light is generated by incident laser light onto a rotating groundglass (RG). The size of laser light spot on RG is controlled by the distance between lens (L) and RG. The intensity of light is controlled by a variable light intensity attenuator (VA). The scattered light is split into two parts by a 1:1 non-polarizing beam splitter (BS). Two fiber couplers (FC$_1$ and FC$_2$) are employed as fiber holders for single-mode fibers with core diameter of 4.0 $\mu m$. The output of two single-mode fibers are sent into two single-photon detectors, respectively. One FC is fixed in space and the other FC is scanable in x-y direction by mounting on motorized translation stages. A Hydraharp 400 module is employed as PTR to record the photon detection time series from these two detectors. For the experiments with laser light, RG is removed.  

\begin{figure}[htb]
\centering
\includegraphics[width=65mm]{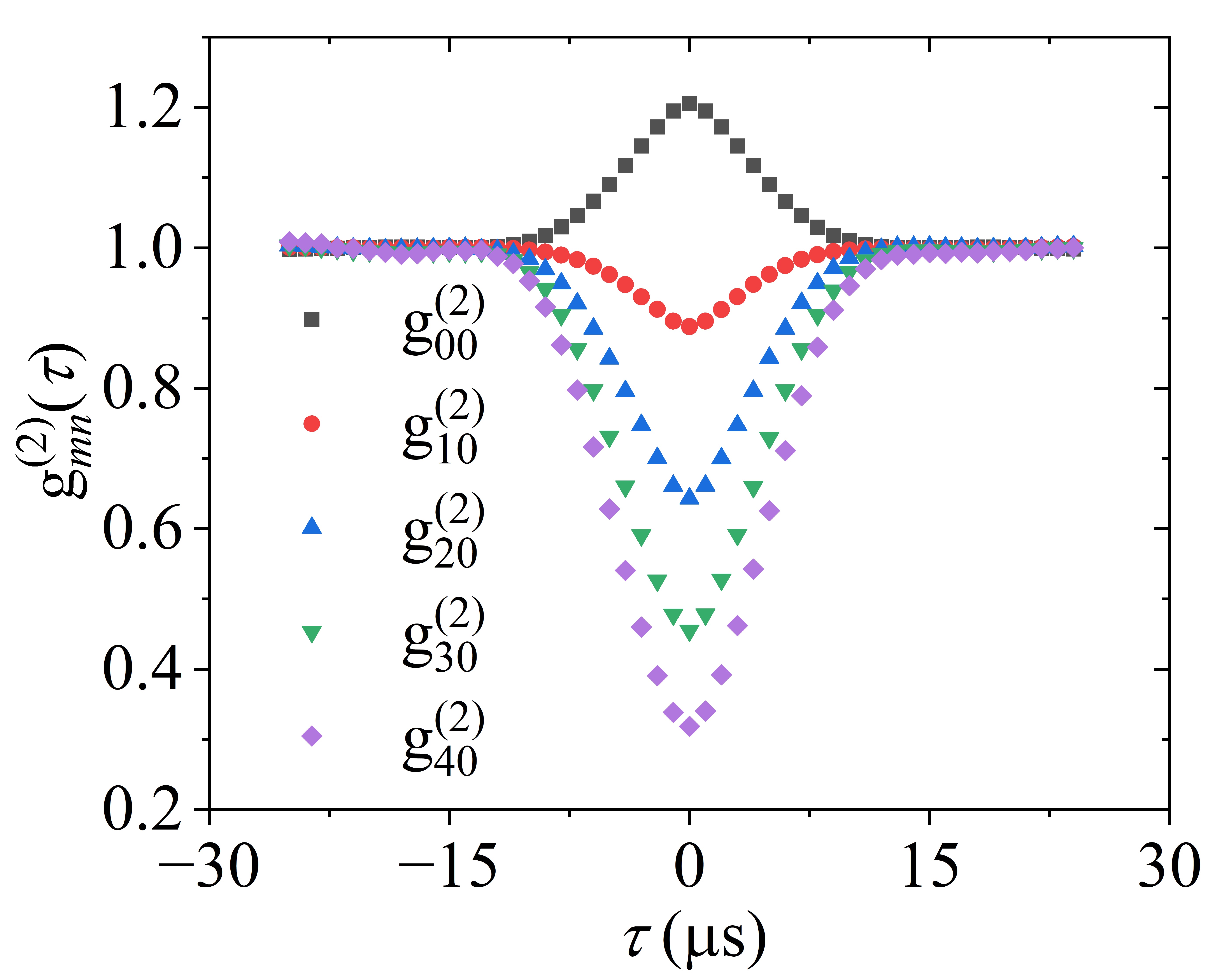}
\caption{Measured of $g^{(2)}_{m0}(\tau)$ of pseudothermal light. The width of time bin, $T$, is 1 $\mu$s. The average number of photons equals 0.66, which is calculated by averaging the photons detected by D$_1$ and D$_2$. }\label{6-gm0}
\end{figure}

Figure \ref{6-gm0} shows the measured $g^{(2)}_{m0}(\tau)$ of pseudothermal light when these two detectors are in the symmetrical positions. The width of time bin is 1 $\mu$s. The single-photon counting rates of D$_1$ and D$_2$ are 734.57 and 583.82 kHz, respectively. The average number of detected photons in one time bin (1 $\mu$s) equal 0.73 and 0.58 for D$_1$ and D$_2$, respectively.  As predicted by the numerical simulation in Fig. \ref{4-gmn}, antibunching is observed for $g^{(2)}_{m0}(\tau)$ when $m$ is larger than 0 in Fig. \ref{6-gm0}.

\begin{figure}[htb]
\centering
\includegraphics[width=65mm]{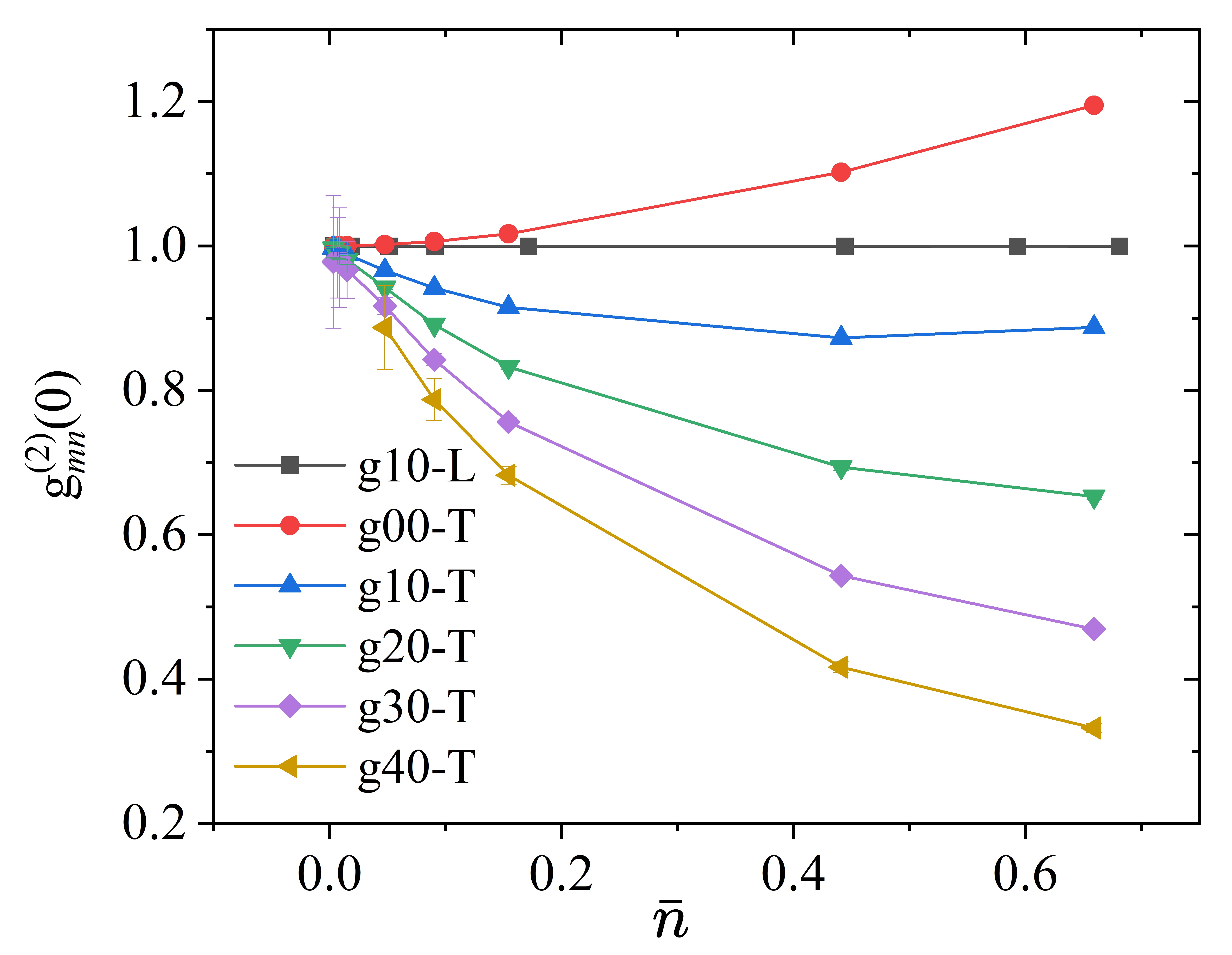}
\caption{Measured $g^{(2)}_{m0}(0)$ for laser (g10-L) and thermal (gm0-T) light with different average number of photons by varying the intensities of input light. $\bar{n}$ is obtained by averaging the number of detected photons of D$_1$ and D$_2$.}\label{7-g-lt}
\end{figure}

Figure \ref{7-g-lt} shows that $g^{(2)}_{m0}(0)$ is dependent on the average number of photons, $\bar{n}$, by changing the intensity of the input light. The black squares labeled g10-L are the measured $g^{(2)}_{10}(0)$ for laser light. The dots labeled g$m$0-T are the measured $g^{(2)}_{m0}(0)$ for thermal light ($m=0$, 1, 2, 3, and 4). The time bin is 1 $\mu$s for all the calculations. The correlation, $g^{(2)}_{m0}(0)$, of single-mode continuous-wave laser light always equals 1 and only the measured results for $g^{(2)}_{10}(0)$ are presented. The value of $g^{(2)}_{m0}(0)$ of thermal light decreases as $m$ increases. Bunching is observed for $g^{(2)}_{00}(0)$. The results shown in Fig. \ref{7-g-lt} are consistent with the theoretical predictions in Figs. \ref{2-g11} - \ref{4-gmn}.

\begin{figure}[htb]
\centering
\includegraphics[width=65mm]{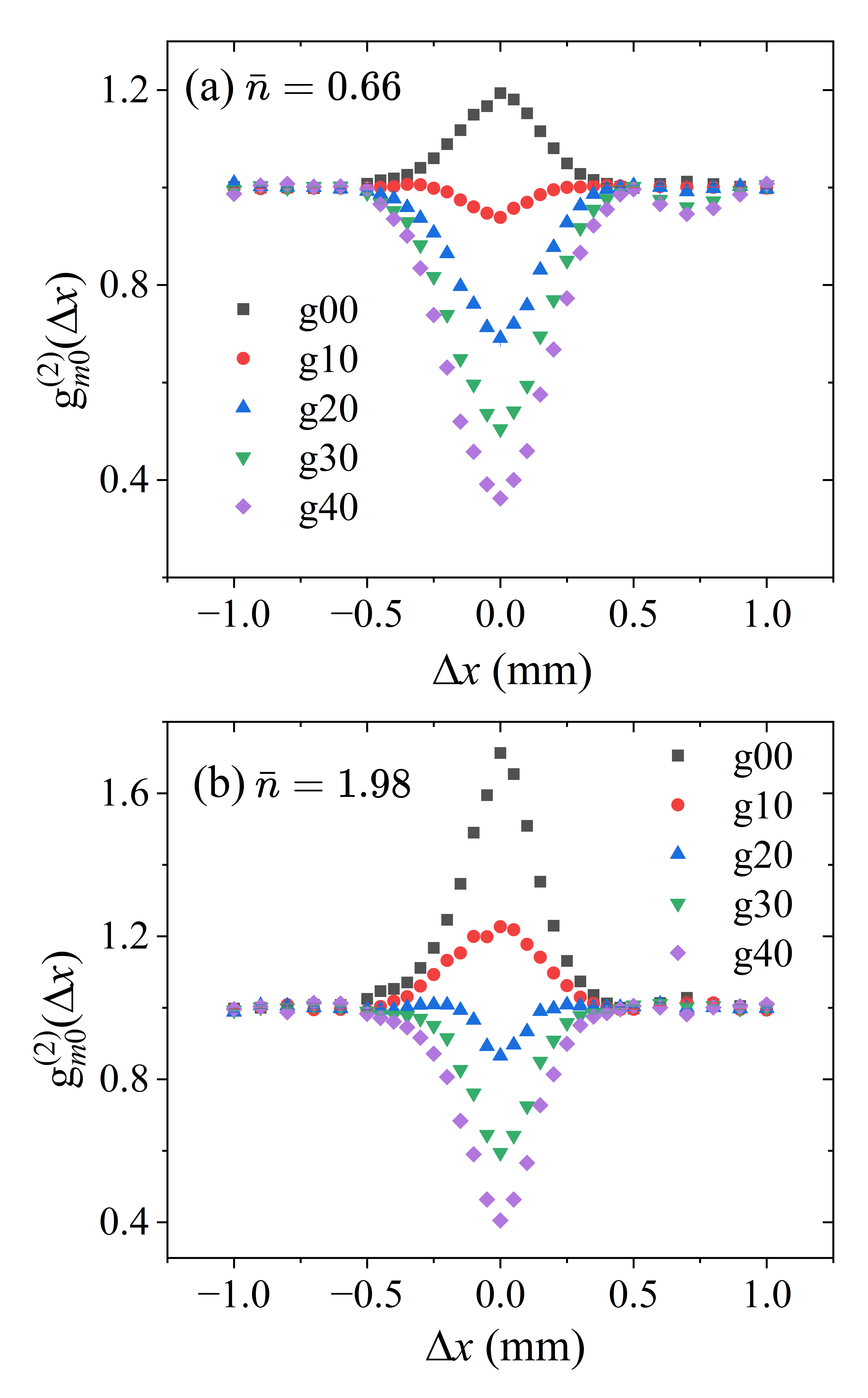}
\caption{Measured $g^{(2)}_{m0}(\Delta x)$ by fixing the position of D$_2$ and scanning the position of D$_1$ horizontally. The value of $g^{(2)}_{m0}(\Delta x)$ for each $\Delta x$ is calculated by dividing the peak (or dip) value by the background. The average number is 0.66 (a) and 1.98 (b).}\label{8-spatial}
\end{figure}

Figure \ref{8-spatial} shows the measured spatial correlation of $g^{(2)}_{m0}(\Delta x)$ by fixing the position of D$_2$ and scanning the position of D$_1$ horizontally. The value of $g^{(2)}_{m0}(\Delta x)$ for each different $\Delta x$ is calculated by dividing the peak (or dip) by the background, respectively. The time bin for the calculation is 1 $\mu$s in Fig. \ref{8-spatial}(a) and 3 $\mu$s in Fig. \ref{8-spatial}(b). The corresponding average number of detected photons equals 0.66 and 1.98 in Figs. \ref{8-spatial}(a) and \ref{8-spatial}(b), respectively, which is calculated by averaging the number of photons detected by D$_1$ and D$_2$. The symbol g$m$0 in Fig. \ref{8-spatial} is short for $g^{(2)}_{m0}(\Delta x)$. The main difference between the results in Figs. \ref{8-spatial}(a) and \ref{8-spatial}(b) is that the observed antibunching via $g^{(2)}_{10}(\Delta x)$ in \ref{8-spatial}(a) is changed into bunching in \ref{8-spatial}(b) due to the average number of detected photons increases. The values of the measured $g^{(2)}_{m0}(0)$ are different in  Figs. \ref{8-spatial}(a) and \ref{8-spatial}(b) for the same $m$, too. 

\section{Discussions}

Based on the numerical and experimental results above, antibunching is observed for thermal light in a linear interferometer provided that single-photon detectors operate as PNRDs to perform photon-number projection measurement. Comparing to traditional HBT interferometer, the experimental setup in Fig. \ref{5-setup} can obtain more information about the correlation of photon detection events. The observed bunching via $g^{(2)}_{00}$ can be understood as a reflection of two-photon bunching of thermal light. When these two detectors are in symmetrical positions, photons tend to come in bunches, which will result time bins with zero photon comes in bunches, too. It is also the reason why antibunching is observed via $g^{(2)}_{10}$ when the average number of detected photons is small. As the value of $m$ increases from 1, $g^{(2)}_{m0}(0)$ can reach 0 more closely, which indicates that multiphoton detection event and zero photon are more antibunched. 

Interestingly, antibunching can be changed into bunching by increasing the average number of detected photons. In the numerical results shown in Fig. \ref{4-gmn}(b2), $g^{(2)}_{10}(0)$ and $g^{(2)}_{20}(0)$ fall below 1 as $\bar{n}$ increases from zero and then increases as $\bar{n}$ continues to increase. $g^{(2)}_{10}(0)$ can be larger than 2 when $\bar{n}$ exceeds 6.  The reason why antibunching can be changed into bunching is a reflection of two-photon bunching of thermal light in different average number of photons.  As shown in Fig. \ref{4-gmn}(b2), the degree of second-order coherence, $g^{(2)}(0)$, of thermal light always equals 2 for different average number of photons. However, things become different for $g^{(2)}_{10}(0)$. When $\bar{n}$ approaches zero, the probability of detecting one photon within a time bin is extremely small. On the other hand, the probability of detecting zero photon within a time bin approaches 1, $P_0 \approx 1$. The probability of D$_1$ detecting one photon and D$_2$ detecting zero photon approximately equals the probability of D$_1$ detecting one photon, $P_{10} \approx P_{1}$. In this condition, the normalized correlation function, $g^{(2)}_{10}(0)=P_{10}/[P_1P_0]$, approximately equals 1. As confirmed in Figs. \ref{4-gmn}(b1) and \ref{4-gmn}(b2),  $g^{(2)}_{10}(0)$ decreases from 1 when $\bar{n}$ increases from zero. 

When $\bar{n}$ is in the neighborhood of  $g^{(2)}_{10}(0)$ getting its minimum, the probability of D$_1$ detecting one photon and D$_2$ detecting zero photon is suppressed due to photons in thermal light tends to come to detectors in bunches, \textit{i.e.}, two-photon bunching of thermal light. It is the reason why antibunching can be observed via $g^{(2)}_{10}$, as shown in Figs. \ref{6-gm0} and \ref{8-spatial}(a). 

When $\bar{n}$ approaches infinity large, the probability of D$_2$ detecting zero photon within a time bin is extremely small. In the time bin when D$_2$ detects zero photon, it means that the intensity of thermal light is very likely at its minimum value during the intensity fluctuation process. The probability of D$_1$ detecting one photon and D$_2$ detecting zero photon does not decreases as fast as the one of D$_1$ detecting one photon. The normalized correlation function,  $g^{(2)}_{10}(0)$ will increase as $\bar{n}$ increases and bunching will be observed. Comparing the observed $g^{(2)}_{10}(\Delta x)$ in Figs. \ref{8-spatial}(a) and \ref{8-spatial}(b), anibunching is indeed changed into bunching by increasing the average number of detected photons.

The observed antibunching via $g^{(2)}_{10}$ is between D$_1$ detecting one photon and D$_2$ detecting zero photon, which is different from the usual antibunching via $g^{(2)}$ \cite{paul1982photon,loudon2000quantum}. The usual antibunching via $g^{(2)}$ can not be interpreted in classical theory. Antibunching via $g^{(2)}_{10}$ can not be interpreted in classical theory, either. However, antibunching via $g^{(2)}_{10}$ can be interpreted in both quantum and semi-classical theory, while anti-bunching via $g^{(2)}$ can only be interpreted in quantum theory. Based on the criteria given by Glauber and Sudarshan \cite{glauber1963quantum,sudarshan1963equivalence}, antibunching via $g^{(2)}$ is a nonclassical effect and there may exist question whether antibunching via $g^{(2)}_{10}$ is nonclassical or not. It is possible to observed nonclassical effect with classical light and photon-number projection measurement \cite{dawkins2024quantum}.  Single-photon detector can be employed as PNRD to perform photon-number projection with resolution of 1. By employing time-multiplexing technique \cite{achilles2004photon}, the  photon number resolution of single-photon detector working as PNRD can be larger than 1. From this point of view, the observed antibunching via $g^{(2)}_{10}$ is a nonclassical effect. However, the observed antibunching via $g^{(2)}_{10}$ can be changed into bunching by increasing the average number of photons, as confirmed in Figs. \ref{8-spatial}(a) and  \ref{8-spatial}(b). If the observed antibunching via $g^{(2)}_{10}$ is nonclassical, what about the nature of the observed bunching via $g^{(2)}_{10}$ in the same experiment. It is interesting to discuss the nature of the observed bunching via $g^{(2)}_{10}$ is classical or nonclassical, which is beyond the scope of this paper.  It is safe to say that the observed antibunching and bunching via $g^{(2)}_{10}$ serves as a bridge to understand the connection between classical and nonclassical effects.

\section{Conclusions}

In conclusion, we have observed spatial and temporal antibunching with classical light in a modified HBT interferometer, in which single-photon detectors operate as photon-number-resolving detectors to perform photon-number projection measurement. The observed antibunching is due to the photon statistics of thermal light and photon-number projection measurement. The observed antibunching can be changed into bunching by varying the average number of detected photons. The results are helpful to understand the relationship between classical and nonclassical correlation and may find potential applications in multiphoton interference and quantum imaging.

\begin{acknowledgments}
This work was supported by the Xi'an Science and Technology Program Project (Grant No. GX2331), Shaanxi Key Research and Development Project (2024CY2-GJHX-89), and National Training Program of lnnovation for Undergraduates.
\end{acknowledgments}

\section*{Appendix A: derivation of $P_{mn}(\vec{r}_1,t_1;\vec{r}_2,t_2)$}\label{appendix-a}
\def\theequation{$A-$\arabic{equation}}
\setcounter{equation}{0}
\def\thefigure{$A-$\arabic{figure}}
\setcounter{figure}{0}

In order to derive $P_{mn}(\vec{r}_1,t_1;\vec{r}_2,t_2)$ based on Eqs. (\ref{probability}) and (\ref{probability-final}), it is helpful to introduce new variables, $X=x-1$ and $Y=y-1$,
so that $x=X+1$, $y=Y+1$. The point $(x,y)=(0,0)$ corresponds to $(X,Y)=(-1,-1)$. Equation (\ref{probability-final}) becomes
\begin{equation}
P_{mn}=\left.\frac{1}{m!\,n!}\frac{\partial^{m+n}M(X,Y)}{\partial X^m\partial Y^n}\right|_{X=-1,Y=-1},
\end{equation}
with
\begin{equation}\label{mxy}
M(X,Y)=\frac{1}{1-\bar{n}(X+Y)+(1-\mu)\bar{n}^2XY}.
\end{equation}

Cauchy integral representation can be employed to extract the coefficient
\begin{eqnarray}\label{pmn-a}
&&P_{mn}\\
&=&\frac{1}{(2\pi i)^2}\oint\oint M(X,Y)\,(X+1)^{-m-1}(Y+1)^{-n-1}dXdY,\nonumber
\end{eqnarray}
where the contours encircle $X=Y=-1$. In order to calculate the integral, setting $u=X+1$, $v=Y+1$ and the integration contours now encircle $u=v=0$. Substituting $u=X+1$ and $v=Y+1$ into the denominator of Eq. (\ref{mxy}) and the denominator can be simplified as
\begin{eqnarray}
&&1-\bar{n}[(u-1)+(v-1)]+(1-\mu)\bar{n}^2(u-1)(v-1)\nonumber\\
&=&\bigl[1+2\bar{n}+(1-\mu)\bar{n}^2\bigr]\nonumber\\
&&-\bigl[\bar{n}+(1-\mu)\bar{n}^2\bigr](u+v)+(1-\mu)\bar{n}^2 uv.
\end{eqnarray}
Define the following constants to simplify the expression
\begin{eqnarray} \label{constants}
C&&=1+2\bar{n}+(1-\mu)\bar{n}^2,\nonumber\\
D&&=\bar{n}+(1-\mu)\bar{n}^2=\bar{n}\bigl[1+(1-\mu)\bar{n}\bigr],\\
E&&=(1-\mu)\bar{n}^2.\nonumber
\end{eqnarray}
Equation (\ref{mxy}) can be written as 
\begin{equation}
M(u,v)=\frac{1}{C-D(u+v)+Euv},
\end{equation}
and Eq. (\ref{pmn-a}) becomes
\begin{eqnarray}\label{pmn-a1}
P_{mn}&=&\frac{1}{C}\cdot\frac{1}{(2\pi i)^2}\oint\oint\frac{1}{1-a(u+v)+buv}\nonumber \\
&&\times u^{-m-1}v^{-n-1}\,du\,dv,
\end{eqnarray}
where $a=D/C$, $b=E/C$ have been assumed.

For $|a(u+v)-buv|<1$ on the contours, 
\begin{equation}
\frac{1}{1-a(u+v)+buv}=\sum_{k=0}^{\infty}\bigl[a(u+v)-buv\bigr]^k.
\end{equation}
Using the binomial theorem,
\begin{eqnarray}
\bigl[a(u+v)-buv\bigr]^k&=&\sum_{j=0}^{k}\binom{k}{j}a^{k-j}(u+v)^{k-j}(-buv)^j, \nonumber \\
(u+v)^{k-j}&=&\sum_{i=0}^{k-j}\binom{k-j}{i}u^{i}v^{k-j-i},
\end{eqnarray}
where $\binom{k}{j}=\frac{k!}{j!(k-j)!}$, $\bigl[a(u+v)-buv\bigr]^k$ can be simplified as
\begin{equation}
\sum_{j=0}^{k}\sum_{i=0}^{k-j}\binom{k}{j}\binom{k-j}{i}a^{k-j}(-b)^j u^{i+j}v^{k-i}.
\end{equation}
The coefficient of $u^{m}v^{n}$ is needed to calculate $P_{mn}$. Therefore,
\begin{eqnarray}
i+j&&=m,\nonumber\\
k-i&&=n.
\end{eqnarray}
Hence $i=k-n$, $j=m+n-k$. The constraints $0\leq j\leq k$ and $0\leq i\leq k-j$ imply that $k\geq \max(m,n)$ and $k\leq m+n$. Also $i\geq 0$ requires $k\geq n$, and $j\geq 0$ requires $k\geq m$. Altogether $k$ runs from $\max(m,n)$ to $m+n$. By re‑indexing $k' = m+n-k$, the summation becomes a sum over $k'=0$ to $\min(m,n)$. With the above substitutions, Eq. (\ref{pmn-a1}) can be simplified as
\begin{equation}\label{pmn-a2}
P_{mn}=\frac{1}{C}\sum_{k=0}^{\min(m,n)}\frac{(m+n-k)!}{(m-k)!\,(n-k)!\,k!}\,a^{m+n-2k}(-b)^k.
\end{equation}
Substituting $a=D/C$ and $b=E/C$ back into Eq. (\ref{pmn-a2}) and $P_{mn}$ can be written as
\begin{equation}
P_{mn}=\sum_{k=0}^{\min(m,n)}\frac{(m+n-k)!}{(m-k)!\,(n-k)!\,k!}\,\frac{D^{\,m+n-2k}(-E)^k}{C^{\,m+n-k+1}}.
\end{equation}

Replacing $C$, $D$, $E$ with the definitions given by Eqs. (\ref{constants}) yields the probability function of D$_1$ detecting $m$ photons and D$_2$ detecting $n$ photons,
\begin{widetext}
\begin{equation}\label{final-a}
P_{mn}(\vec{r}_1,t_1;\vec{r}_2,t_2)=\sum_{k=0}^{\min(m,n)}\frac{(m+n-k)!}{(m-k)!\,(n-k)!\,k!}\,
\frac{\bigl[\bar{n}+(1-\mu)\bar{n}^2\bigr]^{m+n-2k}\bigl[-(1-\mu)\bar{n}^2\bigr]^k}{\bigl[1+2\bar{n}+(1-\mu)\bar{n}^2\bigr]^{m+n-k+1}},
\end{equation}
\end{widetext}
which is valid for all non‑negative integers $m,n$, and $0\leq \mu\leq 1$, $\bar{n}\geq 0$. Thus we have obtained Eq. (\ref{pmn-final}) in the main text of this paper. As an example, two special cases are discussed in the following.

(I) $n=0$ (or $m=0$)

In this case, there is only one term, $k=0$, in the sum over $k$ left. Substituting $n=0$ into Eq. (\ref{final-a}) and it is straightforward to have
\begin{equation}
P_{m0}(\vec{r}_1,t_1;\vec{r}_2,t_2) = \frac{\bigl[\bar{n}+(1-\mu)\bar{n}^2\bigr]^m}{\bigl[1+2\bar{n}+(1-\mu)\bar{n}^2\bigr]^{m+1}}.
\end{equation}
Due to the symmetry of Eq. (\ref{final-a}) in $m$ and $n$, it is easy to have
\begin{equation}
P_{0n}(\vec{r}_1,t_1;\vec{r}_2,t_2) = \frac{\bigl[\bar{n}+(1-\mu)\bar{n}^2\bigr]^n}{\bigl[1+2\bar{n}+(1-\mu)\bar{n}^2\bigr]^{n+1}}.
\end{equation}

(II) $m=n=1$

For $m=n=1$, the sum over $k$ runs from $k=0$ to $k=1$. 

\paragraph{Term $k=0$:}
$
\frac{(1+1-0)!}{(1-0)!(1-0)!0!}=2,\quad D^{2},\quad (-E)^0=1,\quad C^{1+1-0+1}=C^{3},
$
so term $=2D^{2}/C^{3}$.

\paragraph{Term $k=1$:}
$
\frac{(1+1-1)!}{(1-1)!(1-1)!1!}=1,\quad D^{0}=1,\quad (-E)^1=-E,\quad C^{1+1-1+1}=C^{2},
$
so term $=-E/C^{2}$.

Thus
\begin{equation}\label{a-17}
P_{11}= \frac{2D^{2}}{C^{3}}-\frac{E}{C^{2}}.
\end{equation}
Inserting $D=\bar{n}[1+(1-\mu)\bar{n}]$, $E=(1-\mu)\bar{n}^{2}$, $C=1+2\bar{n}+(1-\mu)\bar{n}^{2}$ into Eq. (\ref{a-17}) gives
\begin{equation}
P_{11}= \frac{2\bigl[\bar{n}+(1-\mu)\bar{n}^{2}\bigr]^{2}}{\bigl[1+2\bar{n}+(1-\mu)\bar{n}^{2}\bigr]^{3}}-\frac{(1-\mu)\bar{n}^{2}}{\bigl[1+2\bar{n}+(1-\mu)\bar{n}^{2}\bigr]^{2}}.
\end{equation}
After combining over a common denominator, $P_{11}(\vec{r}_1,t_1;\vec{r}_2,t_2) $ can be simplified as
\begin{equation}
P_{11}(\vec{r}_1,t_1;\vec{r}_2,t_2) = \frac{\bar{n}^{2}\bigl[(1+\mu)+2(1-\mu)\bar{n}+(1-\mu)^{2}\bar{n}^{2}\bigr]}{\bigl[1+2\bar{n}+(1-\mu)\bar{n}^{2}\bigr]^{3}},
\end{equation}
which can be employed to calculate the normalized correlation functions.

\bibliography{antibunching.bib}

\end{document}